\documentclass[12pt]{iopart}
\usepackage{graphicx,epsfig,iopams,multirow,bigdelim,epstopdf}

\begin{document} 
 
\title[Three level Landau-Zener-Coulomb model]{Exact transition probabilities in the three-state Landau-Zener-Coulomb model}

\author{Jeffmin Lin$^{1,2}$ and N A Sinitsyn$^2$}
\address{$^1$Department of Mathematics, Princeton University, Princeton, NJ 08544, USA}
\address{$^2$Theoretical Division,
Los Alamos National Laboratory,
Los Alamos, NM 87545, USA}
\ead{jeffminl@princeton.edu, nsinitsyn@lanl.gov}

\begin{abstract}
We obtain the exact expression for the matrix of nonadiabatic transition probabilities in the model of three interacting states with a time-dependent Hamiltonian. Unlike other known solvable Landau-Zener-like problems, our solution is generally expressed in terms of hypergeometric functions that have relatively complex behavior, e.g. the obtained transition probabilities may show multiple oscillations as functions of parameters of the model Hamiltonian.
\end{abstract}

\maketitle

\section{Introduction}


The study of nonstationary quantum systems is particularly complicated due to the scarcity of the class of models with time-dependent Hamiltonians that can be solved exactly. For example, the stationary problem of the dynamics of a spin-1/2 system in a constant magnetic field can be studied by a trivial diagonalization of a 2$\times$2 Hamiltonian matrix. In contrast, the behavior of this system in an arbitrary time-dependent field cannot generally be obtained in a closed form. Instead, a number of useful exact results have been derived  for simple time-dependence of parameters  in oder to describe two state system behavior in specific but frequently encountered situations \cite{maj,LZ,rozen,nikitin}. Such solvable models have also been used to design widely applicable nonperturbative approximations for the general problem of nonadiabatic transitions between two states \cite{book}.

The nonperturbative behavior of explicitly driven quantum systems with more than two interacting states is usually hard to investigate analytically. Some of the most studied such models address the multistate Landau-Zener (LZ) problem \cite{be}, which is to find transition probabilities among $N$ discrete states induced during the time evolution from either $\tau=-\infty$ to $\tau=+\infty$ or from $\tau=0$ to $\tau=+\infty$ in systems described by the following Sch\"odinger equation with linearly time-dependent coefficients:
\begin{equation}
i\frac{d\psi}{d\tau} = (\hat{A} +\hat{B}\tau)\psi,
\label{mlz}
\end{equation}
where $\hat{A}$ and $\hat{B}$ are constant $N\times N$ matrices. Matrix $\hat{B}$ is diagonal.  
Eigenstates of  $\hat{B}\tau$ are called the {\it diabatic} states and off-diagonal elements of the matrix $\hat{A}$ in the diabatic basis are called the coupling constants.  One has to find the scattering $N\times N$ matrix $\hat{S}$, in which the element $S_{nn'}$ is the amplitude of the diabatic state $n'$ at $\tau \rightarrow +\infty$, given that at $\tau \rightarrow -\infty$ (or at $\tau=0$) the system was at the state $n$.  The related matrix $\hat{P}$, with $P_{nn'}=|S_{nn'}|^2$, is called the matrix of transition probabilities.
A number of exact results for models of the type (\ref{mlz}) have been discovered and explored. Currently, their list includes fully solvable Demkov-Osherov \cite{do} and bow-tie  \cite{bow-tie} modes,  which can be defined for an arbitrary number $N$ of interacting states, models with an infinite number of states \cite{lz-chains}, and models that can be reduced to these three  classes by symmetry transformations \cite{reducible}. Importantly, in all models (\ref{mlz}) with a finite number $N$ of states, there are at least two elements of the transition probability matrix that are exactly known  \cite{be,  mlz-1}.

While accumulated information about the model (\ref{mlz}) is substantial, transition probabilities in all exact results obtained for this class of systems show unusually simple behavior. For example, in all solved models with a finite number of states, transition probabilities can be expressed as simple polynomial forms of $e^{-\pi |A_{ij}|^2/|B_{ii}-B_{jj}|}$. Moreover, all such exact results can be understood in terms of a semiclassical picture, in which the Landau-Zener formula for two linearly crossing levels is applied in a chronological order to each encountered intersection with trivial assumptions about the effect of the phase coherence. Such behavior is generally not observed in numerical simulations of models whose analytical solutions are not known.  For example, multiple oscillations of transition probabilities as functions of parameters were generally found in numerical and perturbative calculations \cite{complexLZ}. This contrast raises questions about how much one can trust exact results, e.g. for developing approximate schemes and intuition about non-integrable systems with nonadiabatic transitions. Recently, multistate Landau-Zener-like models, including exactly solvable ones, have found numerous applications in mesoscopic physics, for example, in dynamic phase transitions \cite{app-bose}, physics of Feshbach resonance \cite{app-bose2}, nanomagnets \cite{app-spin},
and the theory of decoherence \cite{app-exp}, so this question is important to explore.
 
In order to avoid possible deficiencies of the exact solutions of systems (\ref{mlz}), the relevant class of time-dependent Hamiltonians can be extended to search for exact results with more complex behavior. For example, in \cite{ostrovsky, sinitsyn-13prl1}, Hamiltonian operators of the type
\begin{equation}
\hat{H}(\tau) = \hat{A} +\hat{B}\tau + \frac{\hat{C}}{\tau},
\label{ham2}
\end{equation}
where $\hat{A}$, $\hat{B}$ and $\hat{C}$ are constant  Hermitian $N\times N$ matrices, have been explored and two new exactly solvable models with a nonzero matrix $C$ and arbitrary $N$ were identified. Nonlinear time-dependence at coefficients of the matrix $\hat{C}$ allows one to explore the effects of the level curvature in the time-energy space on nonadiabatic transitions. The fact that at $\tau \rightarrow 0_+$ some of the levels have infinite energy does not make the scattering problem ill-defined \cite{ostrovsky} because nonadiabatic transitions terminate at large separation of levels. 
Moreover, an approximation by $\sim 1/\tau$ time-dependence  of a diabatic energy can be used as a better than linear crossing approximation because the former allows one to include the effect of the level curvature \cite{sinitsyn-13prl1}.

 Results in \cite{ostrovsky, sinitsyn-13prl1} showed qualitatively new behavior, in comparison to models (\ref{mlz}), such as the possibility of counter-intuitive transitions and finite survival probability in a state that interacts with large $N$ other states with arbitrary couplings. Still, the obtained transition probabilities could be expressed in terms of elementary functions of model parameters, without showing frequently observed features such as multiple oscillations of transition probabilities as functions of coupling constants.

So far, only special elements of the transition probability matrix have been derived explicitly for the  models in \cite{ostrovsky, sinitsyn-13prl1}. 
Hence, in this article, we work out the three-state version of the model of Ref.~\cite{sinitsyn-13prl1} completely, i.e., we find transition probabilities among all pairs of states, in contrast to only several elements derived previously in \cite{sinitsyn-13prl1}. Generally, the functional form of transition probabilities in our model can be expressed in terms of hypergeometric functions, which produce unusually rich  behavior in comparison to other previously solved three-state Landau-Zener-like systems.   	

\section{The Model}\label{section:description}

Our model can be described by the following Schr\"odinger equation for the amplitudes of three states (a special, $N=3$, case of the Landau-Zener-Coulomb (LZC) model introduced in \cite{sinitsyn-13prl1}):
\begin{equation}
i\frac{d}{d\tau}\left(\begin{array}{ccc}
a(\tau)\\
b_1(\tau)\\
b_2(\tau)
\end{array}\right)
=
\left(\begin{array}{ccc}
k^2/\tau&g_1&g_2\\
g_1&\beta_1\tau&0\\
g_2&0&\beta_2\tau
\end{array}\right)
\left(\begin{array}{ccc}
a(\tau)\\
b_1(\tau)\\
b_2(\tau)
\end{array}\right).
\label{sred} 
\end{equation}
We will call $a(\tau)$ the amplitude of the ``0th level,'' $b_1(\tau)$ the amplitude of the ``1st level,'' and $b_2(\tau)$ the amplitude of the ``2nd level'';  $g_1$ and $g_2$ are called the coupling constants and $\beta_1,\, \beta_2$ are called the slopes of the diabatic energy levels of the $1$st and $2$nd states, respectively. 

\begin{figure}
\caption{\label{figure:energies} The example of time-dependence of the adiabatic energies (eigenspectrum of the Hamiltonian) for a system (\ref{sred}) with $\beta_2>0>\beta_1$.}
\begin{indented}
\item[]\includegraphics[scale=.6]{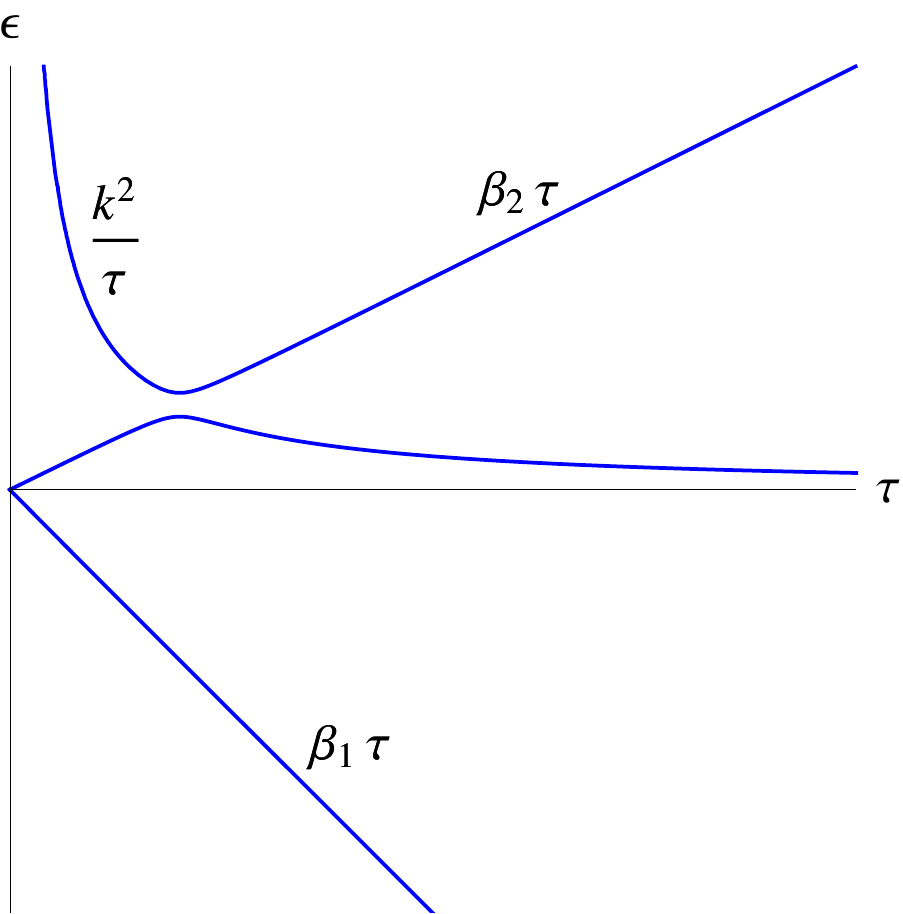}
\end{indented}
\end{figure}

At $\tau \rightarrow 0_+$ the 0th level has ``infinitely" large energy (figure~\ref{figure:energies}). Since only this level is directly coupled to other states, transitions among all states are initially suppressed. At finite time, separation between levels becomes comparable to the coupling between them, so that nonadiabatic transitions become substantial. Eventually, diabatic energies again diverge so that at time $\tau \rightarrow +\infty$ all levels again become well separated and transitions among them terminate. Hence, one can formulate a scattering problem with the goal to find the probabilities $P_{ij}$, where $i,j \in \{0,1,2 \}$, of transitions from the initial state $i$ at $\tau \rightarrow 0_+$ to the final state $j$ at time $\tau \rightarrow + \infty$.

One can also formulate an inverse version of the scattering problem. Namely, one can assume that the evolution starts at $\tau \rightarrow -\infty$ and proceeds up to time $\tau \rightarrow 0_{-}$. In this case levels 1 and 2 converge to zero energy at the end of the evolution.  We will prove, however, that transition probabilities in both cases are related by symmetry, so that it is sufficient to solve the problem for the evolution from  $\tau \rightarrow 0_+$ to $\tau \rightarrow + \infty$. Symmetry arguments also help to determine the number of independent components of the matrix $P_{ij}$, and therefore we will discuss the basic symmetries of the problem in more detail here.

(i) First, we note that one can assume that coupling parameters are generally complex numbers, i.e. the Hamiltonian of the system can be formulated as
\begin{equation}
\hat{H}=
\left(\begin{array}{ccc}
k^2/\tau&g_1&g_2\\
g_1^*&\beta_1\tau&0\\
g_2^*&0&\beta_2\tau
\end{array}\right).
\end{equation}
However, by a trivial gauge transformation, $b_1 \rightarrow b_1 e^{-i {\rm arg}(g_1) }$ and $b_2 \rightarrow b_2 e^{-i {\rm arg}(g_2) }$, coupling constants become real. From this it follows that all transition probabilities $P_{ij}$ depend only on absolute values of $g_1$ and $g_2$. 

(ii) Second, consider the time inversion operation $\tau \rightarrow -\tau$ in (\ref{sred}). It is straightforward to check that this change of variables with redefinition $g_{1,2} \rightarrow - g_{1,2}$ keeps the form of (\ref{sred}) intact. 
The latter operation only changes the phases of coupling constants, hence it does not change the transition probability matrix and hence neither does the operation $\tau \rightarrow -\tau$. As a consequence, there is a constraint on the elements of the evolution operator $\hat{U}(\tau_2|\tau_1)$:
\begin{equation}
|U_{ ji} (+\infty| 0_+) |^2=|U_{ji} (-\infty| 0_{-}) |^2=|U_{ij} (0_{-} |-\infty) |^2,
\label{const1}
\end{equation}
where the last equality follows from the unitarity of the evolution matrix.  Equality (\ref{const1}) means that the transition probability $P_{ji}$ for the evolution from $\tau \rightarrow -\infty$ to $\tau \rightarrow 0_{-}$ is the same as the transition probability $P_{ij}$ for the evolution from  $\tau \rightarrow 0_+$ to $\tau \rightarrow + \infty$. Hence, in the rest of the article we will consider only the scattering problem for $\tau>0$.

(iii) Third, consider the equations for the complex conjugated amplitudes. By complex conjugating Eqs.~(\ref{sred}), we find that up to a constant phase change of $g_{1,2}$ the result coincides with (\ref{sred}) in which parameters are redefined as $(k^2,\beta_1,\beta_2) \rightarrow  (-k^2,-\beta_1,-\beta_2)$. Hence, the latter operation also keeps the transition probability matrix intact.

(iv) Finally, the unitarity constraint is responsible for the general constraint on the elements of the transition probability matrix. Since a unitary evolution, either forward or backward in time, conserves the normalization of a state vector, the sum of elements of the transition probability matrix either in any column or in any row is unity, i.e. $\sum \limits_{i=0,1,2} P_{ij} = \sum \limits_{i=0,1,2} P_{ji} =1 $ for any $j$. Such constraints make only five elements of the transition probability matrix independent of each other.

\section{Derivation of the transition probability matrix}

\begin{figure}
\caption{\label{figure:contours}Below are the relevant branch cuts and contours for integration in the complex plane. Here $\beta_2>0>\beta_1$.}
\begin{indented}
\item[]\includegraphics[scale=.6]{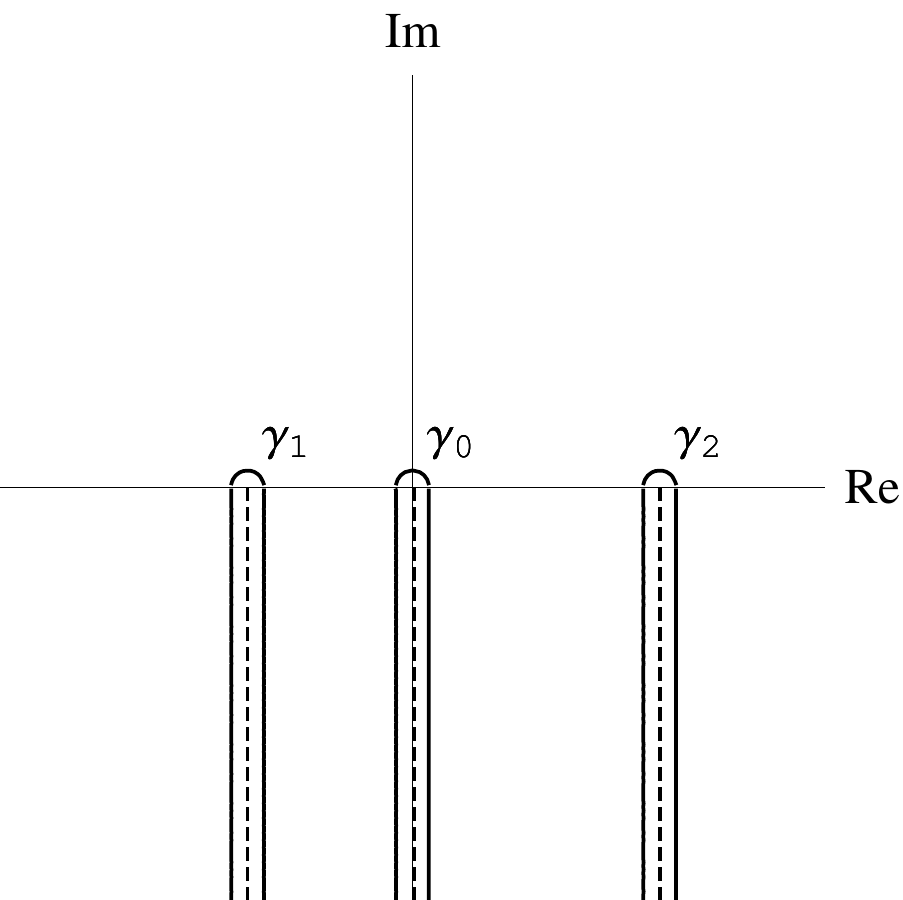}
\end{indented}
\end{figure}

Initially, we follow the steps taken in \cite{sinitsyn-13prl1}: we make a change of variables $a(\tau)=\tau b_0(t)$ and $t=\tau^2/2$, and then use the contour integral ansatz
\begin{equation}
b_j(t)=\int_{\bi{A}}e^{-iut}B_j(u)\ du,\qquad j=0,1,2,
\end{equation}
where $\bi{A}$ is any contour (not closed) for which the integrand vanishes at its limits; integrating by parts and solving for $B_0(u)$ results in a linear first order differential equation for $B_0(u)$. The solutions then take the form of a set of three contour integrals:
\begin{equation}\label{equation:integrals}
\eqalign{b_0(t)=Q\int_{\bi{A}}e^{-iut}(-u)^{\alpha}(-u+\beta_1)^{\xi_1}(-u+\beta_2)^{\xi_2}\ du,\\
b_1(t)=-Qg_1\int_{\bi{A}}e^{-iut}(-u)^{\alpha}(-u+\beta_1)^{-1+\xi_1}(-u+\beta_2)^{\xi_2}\ du,\\
b_2(t)=-Qg_2\int_{\bi{A}}e^{-iut}(-u)^{\alpha}(-u+\beta_1)^{\xi_1}(-u+\beta_2)^{-1+\xi_2}\ du,}
\end{equation}
where
\begin{equation}\label{equation:exponents}
\alpha=-\frac{1}{2}+i\left(\frac{k^2}{2}-\frac{g_1^2}{2\beta_1}-\frac{g_2^2}{2\beta_2}\right),\qquad \xi_1=\frac{ig_1^2}{2\beta_1},\qquad \xi_2=\frac{ig_2^2}{2\beta_2},
\end{equation}
$Q$ is some constant set by the initial conditions, and the complex exponents are understood with branch cuts starting from $0$, $\beta_1$, and $\beta_2$ and going to $-i\infty$, $\beta_1-i\infty$, and $\beta_2-i\infty$ respectively, as shown in \fref{figure:contours}. For future use, we define the contours $\gamma_0$, $\gamma_1$, and $\gamma_2$ as in \cite{sinitsyn-13prl1}: $\gamma_i$ hugs the $i$th branch cut in a counterclockwise fashion (\fref{figure:contours}). For each integral, the particular branch of the integrand that is chosen varies with the sign of the $\beta_i$. 

In \cite{sinitsyn-13prl1}, only asymptotics at $\tau\rightarrow \infty$ of integrals (\ref{equation:integrals}) were needed to determine transition probabilities from the $0$th level and other states.
However, in order to obtain transition probabilities from an arbitrary $m$th level at time $\tau \rightarrow 0_+$ to $n$th level at time $\tau\rightarrow\infty$ one also has to calculate asymptotics of the integrals (\ref{equation:integrals}) at $\tau \rightarrow 0_+$. Here, we evaluate the integrals (\ref{equation:integrals}) in the latter limit and match asymptotics to find the remaining elements of the transition probability matrix, which were not discussed in Ref.~\cite{sinitsyn-13prl1}.

It was noted in \cite{sinitsyn-13prl1} that the integral for $b_i(t)$ as $t\rightarrow\infty$ taken over the contour $\gamma_j$ vanishes unless $i=j$. This means that the contour integrals over $\gamma_j$ provide the solution with the asymptotic that corresponds to only level $j$ populated at time $\tau \rightarrow +\infty$. Consequently, by obtaining the amplitude of $i$th level at $\tau \rightarrow 0_+$ when calculating the integral over $\gamma_j$ one obtains the element of the backward-in-time evolution matrix $(U^{\dagger})_{ij}$. Its absolute value squared is the transition probability from the level $j$ to the level $i$ for the backward-in-time evolution. From the unitarity, this probability coincides with the transition probability {\it from} the level $i$ {\it to} the level $j$ in the needed forward-in-time evolution.

The procedure for finding the transition probability from the $i$th state at $\tau\rightarrow 0_+$ to the $j$th state at $\tau\rightarrow\infty$ is then as follows: first, we pick a contour $A=\gamma_j$, resulting in zero amplitude for $b_n(\tau\rightarrow\infty)$ where $n\neq j$. Second, we find a normalization constant $Q$ necessary to result in a final occupation probability of 1 in the $j$th level (unique up to its argument). This constant can be found directly from the results in \cite{sinitsyn-13prl1}. Third, using this constant $Q$ we calculate the $i$th amplitude at $\tau\rightarrow 0_+$, over the chosen contour $\gamma_j$. Finally, taking the square of the absolute values of the $i$th amplitude at $\tau\rightarrow 0_+$ gives the desired probability. There are a few transition probabilites which we do not present as independent quantities but rather in terms of other transition probabilites. We have not evaluated the contour integral in these cases but rather used the fact that the matrix of transition probabilities is constrained to be doubly stochastic. 
\Table{\label{table:shorthand}Shorthand used in text and \tref{table:matrix}.}
\br
Shorthand&Definition\\
\mr
$\kappa$&$\exp(-\pi k^2)$\\
$q_i$&$g_i^2/\beta_i$\\
$p_i$&$\exp(-\pi g_i^2/|\beta_i|)$\\
$C_i$&$k^2-q_i$\\
$P_{ij}$&Transition probability from level $i$ to level $j$\\
$H_{10}$&$\displaystyle\left|{}_2F_1\left(1-i\frac{g_1^2}{2\beta_1},\frac{1}{2}-i\frac{k^2}{2};1-i\left(\frac{g_1^2}{2\beta_1}+\frac{g_2^2}{2\beta_2}\right);\frac{\beta_2-\beta_1}{\beta_2}\right)\right|^2$\\
$H_{20}$&$\displaystyle\left|{}_2F_1\left(1+i\frac{g_2^2}{2\beta_2},\frac{1}{2}+i\frac{k^2}{2};1+i\left(\frac{g_1^2}{2\beta_1}+\frac{g_2^2}{2\beta_2}\right);\frac{\beta_1-\beta_2}{\beta_1}\right)\right|^2$\\
$H_{12}$&$\displaystyle\left|{}_2F_1\left(\frac{1}{2}-i\left(\frac{k^2}{2}-\frac{g_1^2}{2\beta_1}-\frac{g_2^2}{2\beta_2}\right),\frac{1}{2}-i\frac{k^2}{2};\frac{3}{2}-i\left(\frac{k^2}{2}-\frac{g_2^2}{2\beta_2}\right);\frac{\beta_1}{\beta_1-\beta_2}\right)\right|^2$\\
$H_{21}$&$\displaystyle\left|{}_2F_1\left(\frac{1}{2}+i\left(\frac{k^2}{2}-\frac{g_1^2}{2\beta_1}-\frac{g_2^2}{2\beta_2}\right),\frac{1}{2}+i\frac{k^2}{2};\frac{3}{2}+i\left(\frac{k^2}{2}-\frac{g_1^2}{2\beta_1}\right);\frac{\beta_2}{\beta_2-\beta_1}\right)\right|^2$\\
\br
\endTable

Let us denote the transition probability from the $i$th level to the $j$th level as $P_{ij}$. As an example, consider the transition from the 1st state to the 0th state in the case $\beta_2>\beta_1>0$. In order to obtain $P_{10}$,
we picked the contour $A=\gamma_0$. The normalization constant $|Q|^2$ necessary to result in a final occupation probability of 1 was determined using the results from \cite{sinitsyn-13prl1}. We have then $|b_1|^2=|Q|^2g_1^2|I|^2$, where the normalization constant $|Q|^2$ is explicitly given by
\begin{equation}
|Q|^2=\left(4\pi\left[\exp\left(\pi\left(k^2-\frac{g_1^2}{\beta_1}-\frac{g_2^2}{\beta_2}\right)\right)+1\right]\right)^{-1},
\end{equation}
and the contour integral $I$ is defined as
\begin{equation}
I=\int_{\gamma_0}(-u)^{\alpha}(-u+\beta_1)^{-1+\xi_1}(-u+\beta_2)^{\xi_2}\ du,
\end{equation}
with $\alpha$, $\xi_1$, and $\xi_2$ defined as in \eref{equation:exponents}.
Changing variables to $u=-iz$, letting the branch cut of $z$ go from $-2\pi<\arg z<0$, and splitting the resultant contour into an upper and lower part (above and below the real axis), we obtain
\begin{equation}
I=-(i)^{\alpha+\xi_1+\xi_2}\left(1-e^{-2\pi i\alpha}\right)\int_0^\infty x^\alpha(x-i\beta_1)^{-1+\xi_1}(x-i\beta_2)^{\xi_2}\ dx.
\end{equation}
Evaluating the integral gives a beta function times a hypergeometric function; we then used the branch $(-1)=\exp(-i\pi)$ and $(-i)=\exp(-i\pi/2)$, to evaluate the magnitude squared as
\begin{equation}
|I|^2=\frac{4\pi\left(1-p_1p_2\right)H_{10}}{\beta_2\left(q_1+q_2\right)(1+\kappa)}\left[\exp\left(\pi\left(k^2-\frac{g_1^2}{\beta_1}-\frac{g_2^2}{\beta_2}\right)\right)+1\right],
\end{equation}
with shorthand described in \tref{table:shorthand}. Multiplying by $|Q|^2g_1^2$ gives the desired transition probability $P_{10}$. 

\Table{\label{table:matrix}Matrix of transition probabilites: we present the nonadiabatic transition probabilities, which are the occupation probabilites at $\tau\rightarrow\infty$ with an initial condition of one state occupied with probability 1. Each column corresponds to one of the three possible initial levels. In each matrix, the transition probability to the $n$th level appears in the $n$th row. Therefore, the diagonal elements of the matrices are the survival probabilities. As an example, the probability of a transition from the 2nd level ($b_2(\tau)$) to the 0th level ($a(\tau)$) is found in the upper right entry of each matrix. \Tref{table:shorthand} defines the shorthand used.}
\br
&Initial Probabilites 1,0,0&Initial Probabilites 0,1,0&Initial Probabilites 0,0,1\\
\mr
&\multicolumn{3}{l}{Case 1: $\beta_2>\beta_1>0$}\\
\ldelim({6}{3mm}&$\displaystyle\frac{p_1p_2+\kappa}{1+\kappa}$&$\displaystyle\frac{g_1^2(1-p_1p_2)H_{10}}{\beta_2(q_1+q_2)(1+\kappa)}$&$\displaystyle\frac{g_2^2(1-p_1p_2)H_{20}}{\beta_1(q_1+q_2)(1+\kappa)}$&\rdelim){6}{3mm}\\[2ex]
&$\displaystyle\frac{p_2(1-p_1)}{1+\kappa}$&$1-P_{10}-P_{12}$&$P_{12}+P_{10}-P_{01}$\\[2ex]
&$\displaystyle\frac{1-p_2}{1+\kappa}$&$\displaystyle\frac{g_1^2q_2(p_2+\kappa)H_{12}}{(\beta_2-\beta_1)(1+C_2^2)(1+\kappa)}$&$1-P_{02}-P_{12}$\\[3ex]
&\multicolumn{3}{l}{Case 2: $\beta_2>0>\beta_1$}\\
\ldelim({6}{3mm}&$\displaystyle\frac{p_2+\kappa p_1}{1+\kappa}$&$\displaystyle\frac{g_1^2(p_1-p_2)H_{10}}{\beta_2(q_1+q_2)(1+\kappa)}$&$\displaystyle\frac{g_2^2(p_2-p_1)\kappa H_{20}}{\beta_1(q_1+q_2)(1+\kappa)}$&\rdelim){6}{3mm}\\[2ex]
&$\displaystyle\frac{(1-p_1)\kappa}{1+\kappa}$&$1-P_{10}-P_{12}$&$P_{12}+P_{10}-P_{01}$\\[2ex]
&$\displaystyle\frac{1-p_2}{1+\kappa}$&$\displaystyle\frac{g_1^2q_2(p_2+\kappa)H_{12}}{(\beta_2-\beta_1)(1+C_2^2)(1+\kappa)}$&$1-P_{02}-P_{12}$\\[3ex]
&\multicolumn{3}{l}{Case 3: $0>\beta_2>\beta_1$}\\
\ldelim({6}{3mm}&$\displaystyle\frac{1+\kappa p_1p_2}{1+\kappa}$&$\displaystyle\frac{g_1^2(1-p_1p_2)\kappa H_{10}}{\beta_2(q_1+q_2)(1+\kappa)}$&$\displaystyle\frac{g_2^2(1-p_1p_2)\kappa H_{20}}{\beta_1(q_1+q_2)(1+\kappa)}$&\rdelim){6}{3mm}\\[2ex]
&$\displaystyle\frac{(1-p_1)\kappa}{1+\kappa}$&$1-P_{10}-P_{12}$&$\displaystyle\frac{g_2^2q_1(\kappa p_1+1)H_{21}}{(\beta_1-\beta_2)(1+C_1^2)(1+\kappa)}$\\[2ex]
&$\displaystyle\frac{p_1(1-p_2)\kappa}{1+\kappa}$&$P_{21}+P_{01}-P_{10}$&$1-P_{02}-P_{12}$\\[3ex]
\br
\endTable

\begin{figure}
\begin{indented}
\item[]\includegraphics[scale=.24]{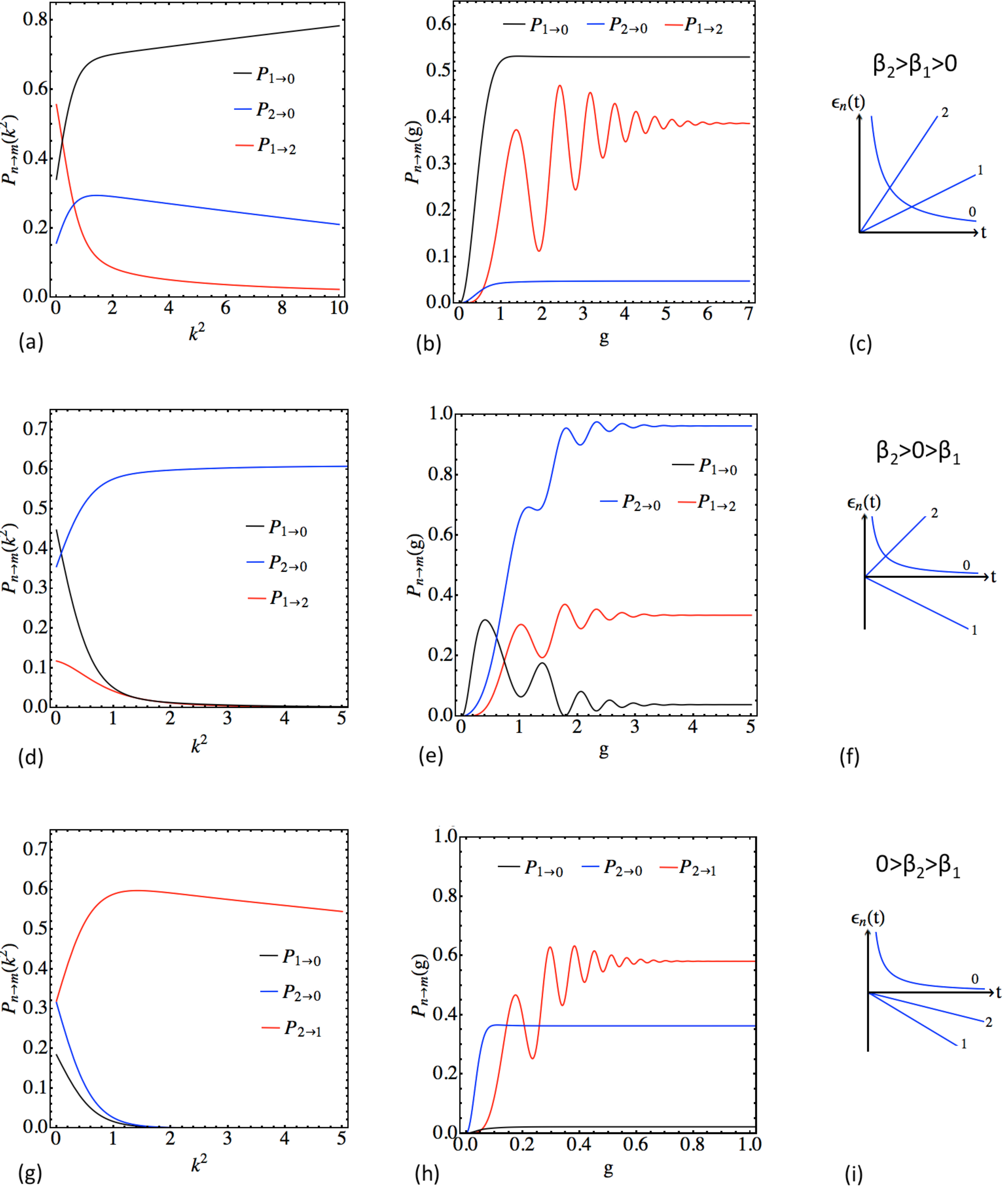}
\end{indented}
\caption{\label{figure:P1n} Transition probabilities as functions of $k^2$ in  (a),(d),(g), and characteristic coupling $g$ in (b),(e),(h). Figures (c), (f), (i) illustrate time-dependence of diabatic energies (diagonal elements of the Hamiltonian) 
for, respectively, figures (a,b), (d,e), and (g,h). Choice of parameters:  (a) $g_1=1,\, g_2=0.7,\, \beta_1=0.9,\, \beta_2=1$;  (b) $g_1=g,\, g_2=0.3g,\, \beta_1=0.9,\, \beta_2=1, \, k^2=0.1$; 
(d) $g_1=0.5,\, g_2=0.3,\, \beta_1=-0.15,\, \beta_2=0.3$;  (e) $g_1=g,\, g_2=0.2g,\, \beta_1=-0.15,\, \beta_2=0.15, \, k^2=0.2$; (g) $g_1=1,\, g_2=1.3,\, \beta_1=-0.3,\, \beta_2=-0.2$; (h) $g_1=g,\, g_2=4.1g,\, \beta_1=-0.15,\, \beta_2=-0.1, \, k^2=0.15$. }
\end{figure}

The probability $P_{12}$ for the two cases $\beta_2>\beta_1>0$ and $\beta_2>0>\beta_1$ was derived similarly. We picked the contour $A=\gamma_2$, then performed a change of variables $u\rightarrow u+\beta_2$, shifting the contour to $\gamma_0$. The remaining steps followed the steps for finding $P_{10}$ very closely.

In the case $0>\beta_2>\beta_1$, instead of finding $P_{12}$, we found $P_{21}$, by letting $k^2$ be negative and calculating the transition probability from a level with slope $|\beta_2|$ to a level with slope $|\beta_1|$, a reflection of the desired model over the time axis and therefore yielding the desired probability. This transition probability followed directly from $P_{12}$ in the case $\beta_2>\beta_1>0$, by sending $\beta_1\rightarrow-\beta_2$, $\beta_2\rightarrow-\beta_1$, $g_1\rightarrow g_2$, $g_2\rightarrow g_1$, and $k^2\rightarrow -k^2$.

We were able to find a concise form for $P_{20}$ as well; for $\beta_1,\beta_2$ of the same sign, our probabilities for $P_{10}$ were valid regardless of the relative magnitudes of the $\{\beta_i\}$. Thus for $\beta_2>\beta_1>0$ and $0>\beta_2>\beta_1$, we simply sent $\beta_1\rightarrow\beta_2$, $\beta_2\rightarrow\beta_1$, $g_1\rightarrow g_2$, and $g_2\rightarrow g_1$. When the $\{\beta_i\}$ are of the same sign, the hypergeometric function may be passed the conjugates of its parameters and will have the same magnitude. This is not true if the $\beta_i$ have different signs. For the case $\beta_2>0>\beta_1$, $P_{20}$ followed from $P_{10}$ in the same case by sending $\beta_1\rightarrow-\beta_2$, $\beta_2\rightarrow-\beta_1$, $g_1\rightarrow g_2$, $g_2\rightarrow g_1$, and $k^2\rightarrow -k^2$.


\section{Results}\label{section:transitions}

The obtained expressions for the transition probabilities between any pair of levels in the three-level LZC model are summarized in Table~\ref{table:matrix}, the notation of which is additionally explained in Table~\ref{table:shorthand}. Furthermore, in Fig.~\ref{figure:P1n}, we show the dependence of three elements of the transition probability matrix on parameters $k$ and $g$, where $g$ is the characteristic coupling such that $g_1=c_1g$ and $g_2=c_2g$ with some constants $c_1$ and $c_2$.

Our solution shows that, in all cases, transition probabilities between levels saturate at values different from $0$ or $1$ in the limit of large coupling constants $g_{1,2}$. This is in sharp contrast with all known exact results for the class of systems (\ref{mlz}). Fig.~\ref{figure:P1n} shows that counterintuitive transitions in the LZC model are generally possible and transition probabilities can have rather complex oscillatory behavior as functions of coupling constants. These are the features that are frequently found in numerical simulations but not in any of the exactly solvable 3-level models of the type (\ref{mlz}). We also note the similarity of qualitative behavior in Fig.~\ref{figure:P1n}(b) and in Fig.~\ref{figure:P1n}(h) despite the former corresponds to two and the latter corresponds to zero number of avoided crossing points. This behavior is explained by smallness of parameter $k$ in both simulations, which corresponds to 
strongly nonlinear time-dependence of the 0th level at the region with nonadiabatic transitions. 
Hence, the presence of crossing points in this regime does not provide a sufficient intuition about the system behavior. However, comparison of Fig.~\ref{figure:P1n}(a) and  Fig.~\ref{figure:P1n}(g) shows that this similarity disappears at larger values of $k$.

\section{Conclusion}\label{section:conclusion}

We obtained exactly all the elements of the transition probability matrix in a three-state model with quantum mechanical nonadiabatic transitions. We found that transition probabilities are generally expressed through the hypergeometric function, for which the asymptotics at limiting values of parameters have been well studied. Fast numerical recipes for calculation of the values of a hypergeometric function are provided by most scientific mathematical software packages.

Our results prove, in particular, that exact solutions can capture nontrivial pattens of  behavior of pairwise transition probabilities, such as multiple oscillations that were observed in numerical simulations of multichannel systems. This richness of the behavior follows from relatively large number of independent parameter combinations that influence transition probabilities. Hence, we demonstrated that the simplicity of known exact results for models in Eq.~(\ref{mlz}) does not directly follows from integrability. It is possible that there is an additional, yet unknown, principle that explains the simple behavior of transition probabilities in these models.

Applications of our solution to specific physical problems are beyond the goal of our work; however, we note that three-state LZC model can be realized in a system of two interacting qubits \cite{sinitsyn-13prl1}, and hence can be used to test the accuracy of control over such basic quantum devices. Our solution can be  also used as a tool to test approximate analytical approaches such as Zhu-Nakamura technique \cite{zhu-nakamura}.  It is interesting to test which of the available approximate schemes that estimate transition probabilities in multichannel models can reproduce the complex behavior of our model, such as the one shown in Fig.~\ref{figure:P1n}(b). 

Finally, we note that at $k=0$ our model coincides with a previously unstudied case of the well known bow-tie model \cite{bow-tie} with evolution from $\tau=0$ to $\tau \rightarrow +\infty$.  It is possible that all solvable Landau-Zener-like models (\ref{mlz}) can be rederived as limiting cases of some more complex exactly solvable systems of the type (\ref{ham2}). Therefore we believe that the class of exact solutions (\ref{ham2}) should be thoroughly investigated.  It is likely that there are other exactly solvable systems of the type (\ref{ham2}), which can shed new light on physics of quantum  non-adiabatic transitions.



\section*{Acknowledgement}
{\it This work was funded by DOE under Contract No.\
  DE-AC52-06NA25396.}

\newpage

\bibliographystyle{iopart-num}
\bibliography{3stateLZCreferences}

\end{document}